\documentclass[preprint]{rspublic}

\usepackage{amsbsy}
\usepackage{amssymb}
\usepackage{amsmath}
\usepackage{natbib}

    \newcommand{\ncd}{\newcommand}
    \ncd{\mrm}    {\mathrm}
    \ncd{\beq} {\begin{equation}}
    \ncd{\eeq} {\end{equation}}

    \def\n{{\rm n}}
    \def\s{{\rm s}}

    \def\d{{\rm d}}
    
    \ncd{\nns}{n_{\n\s}^2}
    \ncd{\wns}{w_a^{\n\s}}
    \ncd{\us}{u_a^\s}
    \ncd{\starq}{\star{\bf q}}

    \ncd{\thetastar}{\theta^*}
    \ncd{\sstar}{s^*}

    \ncd{\un}{u_a^\n}
    \ncd{\Bn}{\mathcal{B}^\n}
    \ncd{\Bs}{\mathcal{B}^\s}
    \ncd{\Ans}{\mathcal{A}^{\n\s}}
    \ncd{\Ann}{\mathcal{A}^{\n\n}}
    \ncd{\Ass}{\mathcal{A}^{\s\s}}
    \ncd{\Xns}{\chi^{\n\s}}
    \ncd{\Asn}{\mathcal{A}^{\s\n}}
    \ncd{\Xsn}{\chi^{\s\n}}
    \ncd{\nn}{\nonumber}

\begin{document}
\title[Thermal Dynamics in General Relativity]{Thermal Dynamics in General Relativity}

\author{C.S. Lopez-Monsalvo and N. Andersson}
\affiliation{School of Mathematics, University of Southampton, UK}

\maketitle

\begin{abstract}{General Relativity; Thermodynamics; Heat conductivity; Dissipation}
We discuss a relativistic model for heat conduction, building on a convective variational approach to multi-fluid systems where the
 entropy is treated as a distinct dynamical entity.
We  demonstrate how this approach leads to a relativistic version of the Cattaneo equation, encoding the finite thermal relaxation time that is required to satisfy causality. 
We also show that the model naturally includes the non-equilibrium Gibbs relation that is a key ingredient in most approaches to extended thermodynamics. 
Focussing on the pure heat conduction problem, we  compare the variational results to the second-order model
developed by Israel and Stewart. The comparison shows that, despite the very different philosophies behind the two approaches, the two models are 
equivalent at first order deviations from thermal equilibrium.  Finally, we complete the picture by working out the non-relativistic 
limit of our results, making contact with recent work in that regime. 
\end{abstract}

\maketitle

\section{Introduction}

Dissipative fluid phenomena represent a number of challenges for relativistic physics. 
The main reason for this is the simple fact that
the classic Navier-Stokes equations, which are not hyperbolic, allow instantaneous signal propagation. This is obviously not allowed in a relativistic description,
where a model must respect causality in order to be considered acceptable. Given the fundamental issues involved, it is not surprising that the problem of relativistic heat conduction continues to attract interest.
The issues considered range from questions like ``Does a moving body feel cold?'' \citep{landsberg} to issues of stability of models for non-equilibrium
thermodynamics \citep{hiscock01,hiscock02,hiscock03,Olson} and whether different descriptions can be distinguished by experiment \citep{Gero,Lindb}. 

In the mainstream general relativity community the debate has, to some extent, been settled since the late 1970s. 
The key contribution was the work of Israel and Stewart, who developed a model analogous to Grad's 
14-moment theory,  firmly grounded in relativistic kinetic theory \citep{Stewart,IS2,IS}. This so-called ``second order'' theory, 
which extends the pioneering ``first order'' 
work of  \citet{eckart03}, has since found applications in a number of contexts. In
particular, in the last few years there has been a resurgence of interest in the model arising from the need to describe
highly relativistic plasmas generated in colliders like RHIC at Brookhaven and the LHC at CERN \citep{elze,muronga}. However, 
despite the obvious successes of the Israel-Stewart model and its various attractive features there are still
dissenting views in the literature, see for example \citet{colin01,colin02}. Particular objections concern the complexity of the model and the large number of 
dissipation coefficients that are needed to complete it. This is, however, a feature that is shared by all models within the 
extended thermodynamics paradigm \citep{joubook} and it is difficult to see how a simpler model can be developed without 
sacrifice of causality or stability. 

Our discussion is motivated by recent efforts to model the dynamics of superfluid neutron stars. This a problem 
that requires a general relativistic description of ``multi-fluid'' dynamics, and where thermal and
dissipative effects are expected to impact on observations. Work in this area is motivated by 
the effort to detect gravitational radiation, and the need to understand various 
oscillation instabilities in rotating neutron stars \citep{nareview}. These instabilities
tend to be counteracted by viscosity and it is obviously important to have a quantitatively accurate description of the involved
mechanisms. Most studies of this problem have been carried out in Newtonian gravity, basically 
because of a feeling that the relativistic problem is too ``complicated''. There is, however, a 
growing body of work building towards realistic, fully relativistic, models. The present analysis should be viewed in that context. 

We are not
providing a truly original view of the heat conduction problem. Rather, we approach the issue within the 
multi-fluid paradigm that has been successfully applied to  superfluid systems. A key ingredient 
in this analysis is the treatment of the  entropy as an additional (massless) ``fluid'' component.
This idea is obviously phenomenological, but we will show that it provides a model with a number of 
attractive features. Most importantly, the formalism is intuitive and readily generalised to more
complex settings. 

We build on the convective variational formulation developed by Carter \citep{carter01,carter02,carter03}, see
\citet{livrev} for an introductory review.  In addition to the intrinsic elegance of an action principle, an 
appealing feature of the variational approach is that once an equation of state is provided the theory provides 
the relation between the various currents and their conjugate momenta (a point that is usually not considered in the context of the heat problem). 
Another key advantage of the variatiational derivation is that incorporating additional fluid components is straightforward.

In Carter's macroscopic model for the heat problem one considers two fluxes, one corresponding to the matter flux 
and one which is associated with the entropy. These two currents are the fundamental fields appearing in the matter 
sector of the Einstein-Hilbert action. The Lagrangian of the theory is a relativistic invariant and hence it should 
depend only on covariant combinations of the two fluxes, which includes the relative flow between them. This encodes the so-called entrainment effect,
which tilts the momenta with respect to the currents when two or more fluids are coupled, and which 
turns out to be a crucial feature of the multi-fluid approach to relativistic heat conduction. In Carter's original work \citep{regular,carter02}, 
the aim was to keep the model as simple as possible by imposing restrictions on the way that the 
currents interact. On the grounds of simplicity Carter ignored the  entrainment. This omission resulted in a truncated model with
severe stability issues \citep{Olson}. Given this problem, it was suggested that Carter's approach does not provide a 
viable alternative to the Israel-Stewart model. This argument is, however, flawed. A detailed comparison 
of the two formalisms  \citep{priou} shows that, at second order in the deviation from equilibrium, the full (entrained) variational approach is essentially equivalent  
to the Israel and Stewart model. 

An important aspect of the present work is that it extends the recent Newtonian model of thermal dynamics discussed by  \citet{nilsclass} to general relativity. 
In both cases entrainment  has a fundamental impact on the dynamics of entropy. In fact, it is an essential ingredient that preserves causality and stability. 
Both discussions lead to a generalization of the Cattaneo equation \citep{cattaneo} and hence a finite speed of propagation of heat. An equivalent, although not identical
equation arises in the Israel and Stewart theory. The difference between the two approaches is in the underlying thermodynamics.  
Basically, the Israel and Stewart formalism is based on the standard equilibrium Gibbs relation, and therefore the thermodynamic quantities take their equilibrium values. 
In contrast, the covariant dynamics of the variational approach leads naturally to an extended Gibbs relation (analogous to that in many models of extended
irreversible thermodynamics), giving the thermodynamic quantities a different meaning. Of course,  the theories are completely equivalent in the case of thermal equilibrium.

The paper is structured as follows. Section \ref{sec.tf} provides the two-fluid variational derivation. The general philosophy that we adopt is that of  \citet{carter03}.
We obtain the equations of motion by imposing conservative constraints on the variations of the Lagrangian density, but we use the explicit freedom 
in the equations of motion to include momentum exchange and entropy production while keeping the energy-momentum tensor divergence-free. 
Once the equations of motion are obtained it is useful to make a  choice of frame to discuss the thermodynamics and the Gibbs relation. 
We discuss this point in detail, arguing why the choice that we make is natural. 
Section \ref{sec.cat} contains the derivation of the relativistic generalization of the Cattaneo equation
that follows once we impose the second law of thermodynamics. The result is then compared to the predictions of the Israel and Stewart model. 
The Newtonian limit of the theory is obtained in Section \ref{sec.newt}, which establishes the close connection to 
the non-relativistic model of  \citet{nilsclass}. 
The paper concludes with a brief discussion of the implications of the results, possible future extensions and applications.

Throughout the paper we use the convention that 4-dimensional spacetime indices are represented by lowercase letters starting from the beginning of the alphabet, $a,b,c,...$
while 3-dimensional spatial indices are lowercase letters $i,j,k,...$. There should be little risk of confusion. We denote the covariant derivative by a semi-colon.

\section{The two-fluid model}
\label{sec.tf}

We consider the problem of heat conduction in general relativity at the macroscopic level. This means that we 
assume that the particle number is large enough that the fluid approximation applies and that there is a well defined matter current, $n^a$.  
Moreover, we adopt the multi-fluid formalism developed by  \citet{carter02} 
and treat the entropy  as an effective fluid with flux $s^a$. This current is in general not aligned with the particle flux. 
The misalignment is associated with the heat flux and leads to entropy production.

For a generic two-fluid system, the starting point is the definition of a relativistic invariant Lagrangian-type \emph{master function} $\Lambda$. 
Assuming that the system is isotropic, we take $\Lambda$ to be a function of the different scalars that can be formed 
by the two fluxes\footnote{It should be noted that we consider the simplest ``convective'' model. The natural way to account for viscosity would be to allow
the master function to depend also on the associated stresses, see \citet{carter03} and \citet{priou}. This model is, however, significantly more complex and we do not consider it here in 
order to keep the discussion clear.}. From $n^a$ and $s^a$ we can form three scalars;
    \begin{align}
    \label{dens1}
    n^2 & = -n_a n^a \ , \\
    s^2 & = -s_a s^a \ , \\
    \label{dens3}
    j^2 & = -n_a s^a \ .
    \end{align}
Hence, we can write the master function as the density for the matter sector of the Einstein-Hilbert action;
    \beq
    \label{EH}
    S_M = \int \d \Omega \Lambda(n,s,j) \ .
    \eeq
An unconstrained variation of $\Lambda$ then leads to
    \beq
    \label{var1}
    \delta \Lambda = \frac{\partial \Lambda}{\partial n}\delta n + \frac{\partial \Lambda}{\partial s} \delta s + \frac{\partial \Lambda}{\partial j} \delta j \ .
    \eeq

Using \eqref{dens1}--\eqref{dens3}, we can change the passive density variations for dynamical variations of the worldlines generated by the fluxes and the metric \citep{livrev}. 
That is, we use
    \begin{align}
    \delta n =&-\frac{1}{2n}[2g_{ab}n^a\delta n^b + n^a n^b \delta g_{ab}] \ , \\
    \delta s =&-\frac{1}{2s}[2g_{ab}s^a\delta s^b + s^a s^b \delta g_{ab}] \ ,\\
    \delta j =& -\frac{1}{2j}[g_{ab}(n^a \delta s^b + s^b \delta n^a) + n^a s^b \delta g_{ab}] \ .
    \end{align}
This means that the variation \eqref{var1} becomes
    \begin{align}
    \label{var2}
    \delta \Lambda =& \left[-2\frac{\partial \Lambda}{\partial n^2}n_a -\frac{\partial \Lambda}{\partial j^2}s_a \right]\delta n^a+\nn\\
                &\left[ -2\frac{\partial \Lambda}{\partial s^2}s_a -\frac{\partial \Lambda}{\partial j^2}n_a \right]\delta s^a+\nn\\
            &\left[-\frac{\partial \Lambda}{\partial n^2}n^an^b - \frac{\partial \Lambda}{\partial s^2}s^a s^b - \frac{\partial \Lambda}{\partial j^2}n^a s^b\right]\delta g_{ab} \ .
    \end{align}

From \eqref{var2} we can read off the conjugate momentum associated with each of the fluxes;
    \begin{align}
    \label{pin}
    \mu_a=\frac{\partial \Lambda}{\partial n^a} = & g_{ab}(\Bn n^b + \Ans s^b) \ , \\
    \label{pis}
    \theta_a=\frac{\partial \Lambda}{\partial s^a} = & g_{ab}(\Bs s^b + \Ans n^b) \ ,
    \end{align}
where we have introduced the coefficients
    \beq
    \label{var.coefs}
    \mathcal{B}^\n\equiv -2 \frac{\partial \Lambda}{\partial n^2}, \quad \mathcal{B}^\s\equiv -2 \frac{\partial \Lambda}{\partial s^2}, \quad \mathcal{A}^{\n\s}\equiv-\frac{\partial \Lambda}{\partial j^2} \ .
    \eeq

The conjugate variables \eqref{pin} and \eqref{pis} demonstrate the fundamental role of the master function \citep{carter02}. 
The distinct roles of the fluxes and their conjugate momenta are often not considered in the fluids literature. A key advantage of the 
variational approach is that the quantities are immediately determined by the form of the master function. Moreover, it is clear that 
 the momenta are not generally 
aligned with the respective currents,  due to the fact that the master function depends on the relative flux. This is an important effect. 
Fundamentally, there is no physical argument to rule out the dependence on $j^2$. In fact, this coupling is associated with the entrainment effect that is known 
to be of central important in other multi-fluid systems. 

In the case of superfluid neutron star cores the entraiment arises due to the strong interaction and couples
the neutron and proton fluxes, see for example \citet{joynt}.  In the present problem, with matter and entropy, we will show that the entrainment is associated with the thermal relaxation of the system.
This is an important effect that must be accounted for.

The energy-momentum tensor is obtained from \eqref{EH} by noting that the displacements of the conserved currents induce 
a variation in the spacetime metric and therefore the variations of the fluxes, $\delta n^a$ and $\delta s^a$, 
are constrained, see \citet{carterq,livrev,prix} for discussion.  The energy-momentum is thus found to be
    \beq
    \label{se-tensor}
    T_a^{\ b} = \mu_a n^b + \theta_a s^b + \Psi \delta_a^{\ b} \ ,
    \eeq
where we define the generalized pressure, $\Psi$, as
    \beq
    \label{psi}
    \Psi = \Lambda -\mu_a n^a - \theta_a s^a \ .
    \eeq

The equations of motion are obtained by requiring that the divergence of the energy-momentum tensor \eqref{se-tensor} vanishes. For an isolated system, we can express this requirement as an equation
of  force balance
    \beq
    \label{fbal}
    T_{a\ ;b}^{\ b} = f^\n_a + f^\s_a = 0 \ ,
    \eeq
where the individual force densities are
    \begin{align}
    \label{fn}
    f^\n_a &=2\mu_{[a;b]}n^b + n^b_{\ ;b}\mu_a \ , \\
    \label{fs}
    f^\s_a &=2\theta_{[a;b]}s^b + s^b_{\ ;b}\theta_a \ .
    \end{align}
The square brackets represent anti-symmetrization in the usual way.

We note that, in order to obtain the energy momentum tensor \eqref{se-tensor} we needed to impose the conservation of the fluxes as constraints on the variation. 
However, the equations of motion, \eqref{fn} and \eqref{fs}, still allow for non-vanishing  production terms. 
If we, for simplicity, consider a single particle species, the matter current is conserved and we have
	\beq
	\label{divn}
	n^a_{\ ;a} = 0 \ .
	\eeq
This removes the second term from the right-hand side of \eqref{fn}. In contrast, the entropy flux is  generally not conserved.
We will have
\beq
\label{divs} 
s^a_{\ ;a} = \Gamma_\s \ge 0 \ , 
\eeq
in accordance 
with the second law of thermodynamics. This suggests that, to make progress, we need to connect the variational results with the relevant 
thermodynamical concepts. In doing this it makes sense to make a specific choice of frame.

\subsection{The matter frame}

The conservation law \eqref{divn} implies that the force $f^\n_a$ is orthogonal to the matter flux, $n^a$, and therefore has only three degrees of freedom. Furthermore, because of the 
force balance \eqref{fbal}, we also have $n^a f^\s_a=0$. This suggests that it is natural to focus on observers moving with the matter frame. We associate the 
matter current with a four-velocity $u^a$ such that
    \beq
    n^a = n u^a,
    \eeq
where $u_a u^a = -1$ and $n$ is the number density measured in this frame. Historically, this is known as the Eckart frame. As will soon become clear, this is the natural frame in the case of a single particle species, essentially
because it simplifies the analysis. More complex settings, e.g. when dealing with additional particle species and reactions, will make the choice of frame less obvious. 
It may well be that the best strategy in such cases is to follow \citet{landau} and work in the centre of mass frame.
Having said that, it is worth noting that even in the more general problem is there a unique frame associated with the entropy/heat flow.
Even though we will not discuss this problem further here,  there are interesting issues that warrant more detailed thinking. 

Having chosen to work in the matter frame, we can decompose the entropy current and its conjugate momentum into parallel  and orthogonal components. The entropy flux is then expressed as
    \beq
    s^a =  s^* (u^a + w^a)
    \eeq
where $w^a$ is the relative velocity between the two fluid frames, and $u^a w_a=0$. Letting $s^a = s u^a_\s$ where $u_\s^a$ is the four-velocity associated with the 
entropy flux, we see that $s^*=s\gamma$ where 
    \beq
  \gamma =   |u^a + w^a| = (1-w^2)^{-1/2} \ , 
    \eeq
is the redshift associated with the relative motion of the two frames. In the following, we will use an asterisk to denote matter frame quantities. 

 Similarly, we can write the thermal momentum as
    \beq
    \theta_a = \thetastar u_a + \theta^\natural w_a = \left(\Bs s^* + \Ans n\right) u_a + \Bs s^* w_a \ ,
    \eeq
where we have made use of \eqref{pis}. From these expressions we readily obtain a measure of the temperature
measured in the matter frame;
\beq
-u^a \theta_a = \thetastar =  \Bs s^* + \Ans n \ .
\label{tpar}\eeq
In essense, this measure respresents the effective mass associated with the entropy component.
We  have also defined
    \beq
    \label{tperp}
    \theta^\natural = \Bs s^* \ .
    \eeq
It is worth noting that, if we ignore the coupling between the fluxes in the master function by taking $\Ans=0$, then 
we have $\thetastar= \theta^\natural$. This particular case was considered by  \citet{regular}, and 
from the analysis of \citet{Olson} we know that it leads to a model that exhibits  instabilities. 
It is useful to keep this in mind during the following developments. As we will see, the main problem 
with Carter's ``regular'' model is that it leaves no freedom to adjust the thermal relaxation timescale.

In order to express the energy-momentum tensor in terms of the matter frame quantities, 
we define the variables $\sigma^a = s^* w^a$ and $p_a = \Bs s^* w_a$. Using the above expressions the stress-energy tensor \eqref{se-tensor} can be written in a more familiar form;
    \beq
    T_{ab} = - \left[\Lambda - p_a \sigma^a\right] u_a u_b + 2 u_{(a}q_{b)} + P_{ab} \ ,
    \eeq
where, making use of the projection orthogonal to the matter flux 
    \beq
    h_{ab} = g_{ab} + u_a u_b \ , 
    \eeq
the heat flux (energy flow relative to the matter) is given by
    \beq
    \label{heat}
    q_a = -h_{ab}u_c T^{bc} = s^* \thetastar w_a \ .
    \eeq
We also have
    \beq
    P_{ab} = h_{ab} \Psi + p_a \sigma_b \ .
    \eeq
It is worth noting at this point that the variational analysis leads to the presence of ``shear'' terms in the energy-momentum tensor. 
Such terms are usually associated with viscous stresses, and it is interesting to note that they arise even though we consider the 
pure heat conduction problem.
Moreover, this exercise shows that the energy density measured in the matter frame can be obtained by a Legendre-type transform
on the master function. That is, we have
    \beq
    \label{rhostar}
    \rho^* = u_a u_b T^{ab} = - \Lambda + p_a \sigma^a \ .
    \eeq
In fact, this relation informs the choice of $\sigma^a$ and $p_a$ as key variables \citep{carter01}.


\subsection{The temperature problem}

Thermodynamic properties such as pressure and temperature are uniquely defined only in equilibrium.
Intuitively this makes sense since, in order to carry out a measurement (of say the temperature), the measuring device must have time to 
reach ``equilibrium'' with the system. The measurement is obviously only meaningful as long as the timescale required to obtain a result is 
shorter than the evolution time for the system. Of course, this does not prevent a generalisation of the various
thermodynamic concepts. The procedure may not be unique, but one should at least require the generalised concepts to be internally consistent
within the chosen extended thermodynamics model. As a useful demonstration of this notion, and the fact that our model satisfies this criterion, we will consider the particular case of the
temperature. 

When the system is out of equilibrium, we can define a number of different ``temperatures''. 
It makes sense to refer to the quantity obtained from \eqref{tpar} as the \emph{dynamical} temperature since it corresponds to the \emph{effective mass} 
of the entropy component. Similarly, the effective mass for the particles, $\mu$, is given by
    \beq
    \label{mu}
    \mu = - u^a \mu_a = \Bn n + \Ans s^* \ .
    \eeq
Let us now show that \eqref{tpar} agrees with the \emph{thermodynamical} temperature that an observer moving with the matter would measure.
Using the standard definition of temperature, we consider the variation of the energy with respect to the entropy in the observer's frame (while keeping the other thermodynamic variables fixed).
To do this, we note that the energy density measured in the matter frame  \eqref{rhostar} is a function of three independent state variables, $\rho^* = \rho^*(n,s^*,p)$.

Determining the energy density directly from \eqref{se-tensor}, we get
    \beq
    \label{rhostar2}
    \rho^* = \mu n + \thetastar s^* - \Psi \ .
    \eeq
Using the definitions \eqref{tpar}, \eqref{tperp} and  \eqref{mu} we can evaluate the generalized pressure \eqref{psi} in the matter frame as a Legendre-type transform of the master function;
    \beq
    \Psi(\mu,\thetastar,p) = \Lambda + \mu n + \thetastar s^* - p\sigma\ ,
\label{Psi}    \eeq
and so the variation of \eqref{rhostar2}, 
    \beq
    \d \rho^* = \mu \d n + \thetastar \d s^* + \sigma \d p \ ,
 \label{envar}   \eeq
shows that the dynamical temperature agrees with the thermodynamical temperature provided we evaluate it in the appropriate frame. 
We also see that, when the system is out of equilibrium, the energy variation \eqref{envar} depends on the heat flux 
(encoded in $\sigma^a$ and $p_a$). 
This \emph{extended} Gibbs relation is similar to that which is postulated in many approaches to extended thermodynamics \citep{joubook}. The main 
difference here is that \eqref{envar} arises naturally from the variational analysis. 

This result is far from trivial. The requirement that the two temperature measures agree 
determines the additional state parameter, $p$, to be held constant in the variation of $\rho^*$. Any other choice of the third
parameter, e.g. $j^2$, will lead to the determined temperatures being different and, hence, the model less consistent. 
Additional evidence that we have identified natural parameters of the non-equilibrium problem is provided by  
the Legendre transformations \eqref{rhostar} and \eqref{Psi}. This important point was originally made by \citet{carter01} 
in a work that is not widely known. 

\subsection{Thermal equilibrium} 

Since the main part of our discussion concerns the dynamics of systems out of equilibrium, and the comparison 
of different possible models, it makes sense to make a few comments on the state of equilibrium. As usual, we
take thermal equilibrium to mean that there is no heat flux. Hence, we have $q^a=0$ and the entropy is carried along with the matter. 
It then makes sense to introduce the equilibrium quantities $\rho$, $s$ and $\theta$, evaluated in the matter frame, in terms of which 
 we recover the standard Gibbs relation;
    \beq
    \d \rho = \mu \d n + \theta \d s \ .
    \eeq

In equilibrium, both $n^a$ and $s^a$ are conserved which means that 
\beq
u^a_{\ ;a} = - { \dot{n} \over n} = - { \dot{s} \over s} \ , 
\eeq
where 
\beq
\dot{n} = u^a n_{;a} \ ,
\eeq
and similar for $\dot{s}$.

It is worth noting that if we introduce  $S=s/n$ then we have
\beq
\dot{S} = 0 \ .
\eeq
That is, the specific entropy remains constant in the matter frame. 

By adding the momentum equations \eqref{fn}--\eqref{fs} we find that
\beq
\left( n\mu + s\theta \right) \dot{u} + h_a^b \left( n \mu_{;b} + s  \theta_{;b} \right) =0 \ .
\label{momtot}\eeq
Making use of the fundamental relation
\beq
P + \rho = n\mu + s \theta \ ,
\eeq
where $P$ is the  equilibrium pressure (to be distinguished from the generalized pressure $\Psi$ that is relevant also
out of equilibrium), and
\beq
\d P = n \d \mu + s \d \theta \ ,
\eeq
we have
\beq
h_a^b \left[ P_{;b} + (P+\rho) \dot{u}_b \right] = 0 \quad \longrightarrow \quad   (P+\rho) \dot{u}^a = - h_a^b  P_{;b}  \ .
\eeq
As expected, we have the usual relation between the acceleration and the pressure gradient.
Finally, the entropy momentum equation can be cast in the form (the easiest way to see this is to set $\Gamma_\s=0$ in the analysis that follows in the next section)
\beq
h^{ab} \left( \nabla_b \theta + \theta \dot{u}_b \right) = 0 \quad \longrightarrow \theta \dot{u}^a = - h_a^b  \theta_{;b}  \ .
\eeq
Comparing these last two expressions we see that we must have
\beq
{ 1 \over P + \rho} h_a^b  P_{;b} = { 1 \over \theta}   h_a^b \theta_{;b}  \ .
\eeq
In other words, temperature changes lead to pressure variations and vice versa.


\section{The relativistic Cattaneo equation}
\label{sec.cat}

It is well-known that the classic thermodynamics description leads to non-causal heat conduction. 
This is obvious if we consider the general solution to the heat equation corresponding to 
given initial data in an unbounded region of space. Due to the parabolic nature of the heat equation, 
any initial profile evolves to predict a non-zero temperature throughout space, even at arbitrarily early times. 
To remedy this problem, and impose a maximal propagation speed for heat, \citet{cattaneo} proposed a modification to Fourier's law such that
the heat flux vector, $\bf{q}$, is related to the temperature gradient according to 
    \beq
    \tau \dot {\bf q} + {\bf q} = -k \nabla T \ .
\label{cat}    \eeq
This model incorporates the (finite) thermal relaxation time $\tau$ expected to arise from the fact that, on the micro-physical scale, heat propagates due to particle collisions.
Cattaneo's equation leads to the temperature satisfying a telegraph-type equation whose hyperbolic nature leads to a finite propagation speed of heat pulses \citep{joubook}.

Since any relativistic model must encode causal heat propagation, one would expect the heat flux to be described by 
an equation similar to \eqref{cat}. The anticipated form for the relativistic analogue of Cattaneo's equation is, indeed, 
generally agreed upon but its derivation and the physical interpretation of the involved variables differ 
among proposed models. Our main aim  is to obtain the relevant heat conduction equation within the variational approach, 
and compare the result to the second order model of \citet{IS}. In doing this, we will highlight the differences
between the two models. 

\subsection{The variational approach}

We want to formulate a relativistic analogue of Cattaneo's equation. The basic strategy will be to  use 
the orthogonality of the entropy force density $f_\s^a$ with the 
matter flux,  solve for the entropy production rate $\Gamma_\s$  and finally impose the second law of thermodynamics.

Let us first note that, by making use of \eqref{heat}  we can express the entropy flux in terms of the matter flow and the
heat flux;
    \beq
    \label{sdecomp}
    s^a= s^*u^a + \frac{1}{\thetastar}q^a \ .
    \eeq
Moreover,  the conjugate momentum becomes
    \beq
    \label{thetadec}
    \theta_a = \thetastar u_a + \beta q_a \ ,
    \eeq
where $\beta$ is given by
    \beq
    \label{beta}
    \beta = \left(\frac{1}{s^*} + \frac{\theta^\natural - \thetastar}{s^* \thetastar} \right) = \left(\frac{1}{s^*} -  \frac{\Ans n }{s^* \thetastar} \right) \ . 
    \eeq
It is worth noting that if we set $\Ans=0$, as in Carter's ``regular'' model \citep{regular}, then $\beta$ takes the specific value $1/s^*$. 

In terms of the variables we have introduced we have
    \beq
  \thetastar  s^a_{\ ;a} = 2\theta_{[a;b]}u^a q^b \frac{1}{\thetastar} \ ,
    \eeq
which leads to the entropy creation rate
    \beq
\Gamma_\s=   s^a_{\ ;a} = -\frac{1}{{\thetastar}^2}q^a \left(\thetastar_{;a} + \thetastar \dot u_a - \beta q_{c;a}u^c  + \beta  \dot q_a + \beta_{;c} u^c q_a \right) \ .
  \label{Gams}  \eeq
Dots represent time derivatives in the matter frame, as before.
That is, $\dot{u}^a$ 
is the four-acceleration and
\beq
\dot{q}^a = u^b  q^a_{\ ;b} \ , 
\eeq
represents the time variation of the heat flux. 

The expression given in \eqref{Gams} has to satisfy the second law of thermodynamics.
The simplest way to achieve this is to demand that the entropy production is a quadratic in the sources. This suggest that the heat flux takes the form
    \beq
    \label{precat}
    q^a = -\kappa h^{ab}\left( \thetastar_{\ ;b} + \thetastar \dot u_b + 2\beta q_{[b;c]}u^c + \dot \beta q_b \right)
    \eeq
where $\kappa>0$ is the thermal conductivity of the fluid.
This leads to  a relativistic analogue to Cattaneo's equation,  which can be written 
    \beq
    \label{cattaneo}
    2\tau q^{[a;c]}u_c + q^a = -\tilde\kappa h^{ab}\left(\thetastar_{\ ;b} + \thetastar \dot u_b\right) \ ,
    \eeq
or
    \beq
    \label{gato}
    \tau \left( \dot{q}^a + q_c u^{c;a} \right) + q^a = -\tilde\kappa h^{ab}\left(\thetastar_{\ ;b} + \thetastar \dot u_b\right) \ ,
    \eeq
where we have used the fact that $u_c q^{c;a} = - q_c u^{c;a}$. Moreover, we have introduced
the effective thermal conductivity $ \tilde \kappa$;
    \beq
    \tilde\kappa \equiv \frac{\kappa}{1 + \kappa \dot \beta} \ ,
    \eeq
and the thermal relaxation time 
    \beq
    \tau = \frac{\kappa\beta}{1 + \kappa \dot \beta} \ .
    \eeq
It is worth noting that, if $\beta$ varies on a timescale $\tau_\beta$ (say) which is long compared to the 
thermal relaxation, then 
\beq
\kappa \dot{\beta} \approx { \kappa \beta \over \tau_\beta} \approx {\tau \over \tau_\beta} \ll 1 \ . 
\eeq
One would expect this to be the case in most situations of practical interest. Thus, we would simply have
\beq
\tilde{\kappa} \approx \kappa \ , \qquad \mbox{ and } \qquad \tau \approx \kappa \beta \ .
\eeq

\subsection{A different perspective}

Before we proceed to compare \eqref{cattaneo} to the corresponding equation obtained within the \citet{IS} approach , 
it is worth  considering the variational result from a slightly different point of view. 
As in the Newtonian problem \citep{nilsclass}, the entropy momentum equation provides an evolution equation for the heat flux. 
This equation can be written in a number of ways, some of which are helpful in interpreting the results.

As a first step, we recall that  $f^\s_a$ is orthogonal to $u^a$. By contracting \eqref{fn} with $u^a$ and using the result
in \eqref{fs} we arrive at an elegant expression for the force $f^\s_a$;
\beq
-\theta^* f^\s_a = 2 u^c s_b \left( \theta_{[c} \theta_{a]}^{\ ;b} +  \theta_{b[;c} \theta_{a]} \right)  
\label{cato1}\eeq 
Written in this form, the force is clearly orthogonal to $u^a$. Moreover, this expression emphasises the relevance of the entropy momentum 
$\theta_a$. However, if we want to gain insight into the key factors that contribute to the force then we need to expand this expression. 
To do this, we 
contract equation \eqref{fs} with \eqref{sdecomp} to get
    \beq
    \label{qdec}
    (s^a \theta_a)\Gamma_\s = s^af_a^\s = \frac{1}{\thetastar}q^a f_a^\s  \ .
    \eeq
This shows that the entropy production only depends on the piece of the force $f^\s_a$ which is parallel to the heat flux. In general, we can decompose the entropy force into two terms;
    \beq
    f_a^\s = f^\parallel q_a + f_a^\perp \ ,
    \eeq
where both pieces are orthogonal to $u^a$, and $f^\perp_a$ is also orthogonal to $q^a$. From \eqref{qdec} it is obvious that $f^\perp_a$ cannot contribute to the entropy production. 
Hence,  this term is not constrained by the second law. This is an important point because there is no obvious way to 
distinguish the viability of  models with different forms of $f^\perp_a$. Given this, it is interesting to consider the specific force terms that arise in the 
variational formalism. 

With the definitions above it is straightforward to show that
    \beq
    f^\parallel = -\frac{1}{\thetastar}\left(\frac{\beta}{\thetastar} - \frac{s^* \thetastar}{q^2}\right) q^b \left( \thetastar_{\ ;b} + \thetastar \dot u_b + \beta_{;c}u^c q_b + 2  \beta q_{[b;c]}u^c\right) \ .
 \label{fpar}   \eeq
Meanwhile, from \eqref{qdec} we have
    \beq
    \Gamma_\s = \frac{1}{s^a \theta_a} \frac{q^2}{\theta^*}f^\parallel \ ,
    \eeq
where
    \beq
    s^a \theta_a = \sstar \thetastar + \frac{\beta}{\thetastar}q^2 \ .
    \eeq
As expected, this takes us back to  \eqref{Gams}.
Moving on to the non-entropy producing part of the force, the Ansatz \eqref{precat} together with a projection orthogonal to $q^a$ leads to
    \begin{multline}
    f^\perp_a = - { 2  \over s^* \theta^*} h_a^c q^b \left[ \left( \beta q_b \right)_{;c} -   \left( \beta q_c \right)_{;b} \right] \\
=  - { 2  \over s^* \theta^*} h_a^c \perp_c^b \left[ q^2 \beta_{;b} + { 1\over 2} \beta \left( q^2 \right)_{;b} - \beta q^d  q_{b;d}  \right] \ ,
 \label{fperp}   \end{multline}
where 
\beq
\perp_c^b = \delta_c^b - {q_c q^b \over q^2} \ ,
\eeq
projects out the component that is orthogonal to $q^a$. We see  that the variational approach  leads to the presence of terms that, even though they involve the heat flux,
 are not associated with entropy production. As far as we are aware, the dynamical role of these terms has not been discussed in detail in the literature
even though similar terms are (as we will soon see) also included in the Israel-Stewart formalism. The variational model leads to these terms taking a 
specific form. In particular, the terms in \eqref{fperp} are all quadratic in $q^a$, the deviation from equilibrium. At this order, the most general case would allow a force of form
\beq
    f^\perp_a =  h_a^c \perp_c^b \left( A_a q^2 + B \left( q^2 \right)_{;a} + C q^b  q_{a;b} \right) \ ,
\eeq
with $A_a$, $B$ and $C$ unspecified coefficients. There may also, in principle, be first order terms. Clearly, \eqref{fperp} 
represents a particular case where all the coefficients follow from $\beta$. Hence, the variational model is a particular example 
of the general class of permissible theories. The fact that all these models satisfy the second law of thermodynamics means that we
cannot express  a preference at this point. A very interesting question  concerns whether there are 
situations where $f^\perp_a$ has a distinguishable effect on the dynamics of the system. If one could show that this is the case, then we 
may be able to  narrow down the possibilities. 


\subsection{Remarks on the role of the four acceleration}

It is worth commenting on the presence of the term associated with the four-acceleration on the right-hand side of \eqref{gato}. As we will see later, 
this term has no counterpart in the Newtonian problem. However, its presence in the relativistic heat equation  has been known since the pioneering work 
of \citet{eckart03}. Formally, this term originates from the local energy balance, \eqref{n59}. Physically, it results from the fact that the 
infinitesimal 3-spaces orthogonal to the matter world lines are not parallel, but relatively tipped over because of the curvature of the  world line.
This leads to the interpretation of the four-acceleration contribution in terms of the effective inertia of heat \citep{ehlers}. 
Interestingly,  \citet{colin01,colin02} have recently suggested that this term may be the origin of instabilites. We will not discuss this suggestion in detail, 
but note that there are varying points of view in this area of research. Most researchers seem to accept that the four acceleration term is both inevitable
and physically meaningful.

The derivation of \eqref{cattaneo} was based on the thermal momentum equation \eqref{fs}. We could, in principle, also make use of 
the other momentum equation, \eqref{fn}. Contracting this equation with $s^a$ we have
\beq
- q^2 f^\parallel = n q^a \left[ \mu \dot{u}_a +  \mu_{;a} + q_a \dot{\alpha} + {2 \alpha \over n} u^b q_{[a;b]} \right]  \ ,
\eeq
where 
\beq
\alpha = { 1 - \beta s^* \over n} = {\Ans \over \theta^*} \ .
\eeq
This  gives us an expression for $q^a \dot{u}_a$ which could be used in \eqref{fpar}.
We would then arrive at a different form for $\Gamma_\s$ and as a result, the Cattaneo 
equation will also be different. The four acceleration term will now be replaced by the 
chemical potential gradient. After some algebra, we arrive at 
\begin{multline}
\Gamma_\s = - \left( \rho^* + \Psi - {\beta q^2 \over \theta^*} \right)^{-1} {n \mu \over \theta^*} \\
\times q^b \left\{  \theta^*_{;b} +  {\theta^* \over \mu} \mu_{;b} + 
\left( \dot{\beta} + {\theta^* \over \mu} \dot{\alpha} \right) q_b + 2 \left( \beta -{\theta^* \over \mu} \alpha\right)  u^c q_{[c;b]}\right\}  \ .
\end{multline}
We could  impose the second law on this result, and hence derive an alternative form for the Cattaneo equation. 
There is, however, no obvious reason why this form would be more useful than \eqref{cattaneo}. Hence, we will not pursue this
 strategy further here.

\subsection{The Israel-Stewart approach}

The most successful approach to the problem of causal heat conductivity and
dissipation in relativistic fluid dynamics is due to Israel and Stewart \citep{Stewart, IS2, IS}. 
A detailed comparison between their results and the variational model has already been carried out by \citet{priou}. 
The key results of this comparion are: i) The two models differ only at second order in the deviation from 
equilibrium, ii) The inclusion of entrainment, $\Ans\neq0$, is essential in the variational analysis. iii)
The two models belong to a larger class on non-equilibrium thermodynamics models. We will demonstrate these results
in the particular case of pure heat conductivity.  Priou's analysis 
includes the viscosity contributions from \citet{carter03}, which makes many of the equations rather complex. The  message 
is clearer if we focus on the heat conductivity problem, and we feel that it is important to make
the comparison as transparent as possible. 

In order to effect the comparison, it makes sense to begin by working through the
derivation of the Israel-Stewart model for a heat conducting system. We focus on the phenomenological 
 description, and simply note that the model  has a firm foundation in relativistic kinetic theory. 
Hence, we take as the starting point  the stress-energy tensor \citep{hiscock01};
    \beq
    \label{stress}
    T^{ab} = \rho u^a u^b + (P + \tau)h^{ab} + 2 u^{(a}q^{b)} + \tau^{ab}
    \eeq
The main difference from the variational model is that the thermodynamical quantities refer to an equilibrium state.
As before, $\rho$ is the energy density, $P$ is the pressure and $q^a$ is the heat flow orthogonal to the matter flow. 
Meanwhile, $\tau^{ab}$ and $\tau$ are the stresses caused by viscosity in the fluid. The tensor $\tau^{ab}$ satisfies the relations
    \beq
    \label{ort1}
    0 = u^a \tau_{ab}= \tau^a_{\ a}=\tau_{[ab]} \ .
    \eeq

Motivated by kinetic theory, Israel and Stewart expands the entropy flux to include a complete set of second order terms
    \beq
    \label{sisrael}
    s^a_{\small I} = s n^a +\frac{q^a}{T} - \frac{1}{2}\left(\beta_0 \tau^2 + \beta_1 q^b q_b + \beta_2 \tau_{bc}\tau^{bc} \right)\frac{u^a}{T} + \alpha_0 \frac{\tau q^a}{T} + \alpha_1\frac{\tau^a_{\ b} q^b}{T} \ .
    \eeq
Here, $T$ is the absolute temperature associated with the equilibrium state and the  coefficients $\beta_0$, $\beta_1$, $\beta_2$, $\alpha_0$ and $\alpha_1$ correspond to different 
couplings that need to be provided (i.e. obtained from the microphysics of the problem). By neglecting all 
viscosity contributions we are left only with the $\beta_1$ term, and the  entropy flux \eqref{sisrael} reduces to
    \beq
    \label{sisr02}
    s^a_{\small I} = s n^a + \frac{1}{T}q^a - \frac{\beta_1}{2T}q^2 u^a.
    \eeq

From this definition it is straightforward to calculate the divergence of $s^a_{\small I}$ to impose the second law. From the equations of motion we also have
    \beq
    \label{n59}
    u_b T^{ab}_{\ \ ;a} = - \dot \rho - \frac{1}{3} (\rho + P) u^a_{\ ;a} - q^a_{\ ;a} + \dot q_b u^b = 0 \ .
    \eeq

We now use the fundamental relation of thermodynamics in the equilibrium form
    \beq
    P + \rho = \mu n + s T \ ,
    \eeq
together with the equilibrium equation of state, $\rho = \rho(n,s)$, which leads to
    \beq
    \rho_{;a} = \mu n_{;a} + T s_{;a} \ .
    \eeq
We also need to use the fact that the flux $n^a$ is conserved. 

Combining these results, we arrive that
    \beq
    s^a_{I;a} = -\frac{1}{T^2} q^b \left[T_{;b} + T \dot u_b +  T \beta_1 \dot q_b + \left(\frac{\beta_1}{2T}u^a\right)_{;a}T^2 q_b \right] \ .
    \eeq
As before, we need to ensure that the entropy production is positive. This follows as long as 
    \beq
    \label{iscat}
    q^a = - \hat{\kappa} T h^{ab}\left[\frac{1}{T} T_{;b} + \dot u_b +  \beta_1 \dot q_b + \left(\frac{\beta_1 u^c}{T} \right)_{;c}\frac{T}{2}q_b  \right] \ .
    \eeq
This relation provides the Israel-Stewart version of the relativistic Cattaneo equation. 

As in the variational case, one may add other terms to the heat flux as long as they do not lead to the generation of entropy. As an example, \citet{hiscock01} included the term
    \beq
    -\gamma_2 \hat{\kappa} T h^{ab}q^c u_{[c;b]} \ , 
    \eeq
the presence of which is motivated by kinetic theory. This term is clearly
orthogonal to both $q^a$ and $u^a$ and, hence, does not affect $\Gamma_\s$. Technically, it is also a second order
term because the fluid shear is associated with the viscosity.


\subsection{The Extended Irreversible Thermodynamics view}

We have  reached the point where we can compare the final
equations for the heat flux from the variational approach, Eq.~\eqref{cattaneo}, to  the
Israel-Stewart model, Eq.~\eqref{iscat}.  However, before we do this, let us consider the
problem from the ``extended irreversible thermodynamics''
point of view. This approach, which was first developed by \cite{carter01}, provides an immediate illustration of the fundamental difference
between the two models that we are discussing.

As in the variational analysis, we take the heat flux to be given by
    \beq
    s^a = s^* u^a + \frac{1}{\theta^*} q^a \ .
    \eeq
The corresponding stress-energy tensor is 
    \beq
    T^{ab}= \rho^* u^a u^b + 2u^{(a}q^{b)} + \Psi h^{ab} + \sigma^a p^a \ ,
    \eeq
where $\Psi$ denotes the (generalised) pressure, and the variation of the energy density (in the matter frame) is
    \beq
    \d \rho^* = \mu \d n + \theta^* \d s^* + \sigma \d p \ .
    \eeq
Obviously, we are now refering to out-of-equilibrium quantities. Hence,  we use an \emph{extended} Gibbs relation that accounts for the heat flux. 
This immediately leads to
    \beq
    \rho^*_{\ ;a} = \mu n_{;a} + \theta^* s^*_{\ ;a} +  \frac{q}{\theta^*} (\beta q)_{;a} \ .
    \eeq

Using these relations and the particle conservation, we find that
    \beq
    u_bT^{ab}_{\ \ ;a} = (n\mu - \rho^* - \Psi)\Theta - \theta^* \dot s^* - \sigma \dot p + q_{b;a}u^bu^a - q^a_{\ ;a} + p^b_{\ ;a}\sigma^a u_b \ .
    \eeq
At this point we need the fundamental relation \eqref{rhostar2}, which
allows us to complete the derivation. We arrive at
    \beq
    s^a_{\ ;a} = -\frac{1}{{\theta^*}^2}q^b\left[\theta^*_{\ ;b} + \theta^* \dot u_b - \dot q_b\beta +(\beta q_b)_{;c}u^c \right] \ ,
    \eeq
which is identical to \eqref{Gams}. Hence, the heat flux will (again) be governed by \eqref{cattaneo}. 


\subsection{Comparing the results}

Since the two approaches are based on different strategies, any comparison between the
variational model and the Israel-Stewart results must be done carefully. 
Notably, the Israel-Stewart model is based on an expansion including terms up to second order in the 
deviation from equilibrium. Meanwhile, the variational analysis did not involve such an expansion. As a result, 
the final equation for the heat flux, \eqref{cattaneo}, contains higher order terms while \eqref{iscat} is manifestly 
linear in $q^a$.   

In order to compare the results, it makes sense to focus on the linear deviation from equilibrium. Then we have
\beq
\theta^* = T + O\left( q^2 \right) \ , 
\eeq
and it follows that \eqref{gato} can be  approximated by
    \beq
    \tau \left( \dot{q}^a + q_c u^{c;a} \right) + q^a \approx -\tilde\kappa h^{ab}\left(T_{\ ;b} + T\dot u_b\right) \ .
    \eeq
Here it is worth noting that in the second term in the first bracket we could use the standard decomposition of $u_{a;b}$ in  terms of the shear, expansion etcetera.
This would lead to terms that were explicity excluded from the Israel-Stewart model  at the point where we focussed on the case with $\tau_{ab}=\tau =0$. 
Basically, the variational analysis leads to the presence of terms that couple the heat flux to the shear and expansion of the flow. 
As these were artificially excluded from the analysis that led to \eqref{iscat} 
we cannot count this as a difference between the two models. In fact, the full comparison carried out by Priou 
show that these terms agree in the two descriptions. 

Keeping terms up to second order (treating the shear and the divergence of $u^a$ as first-order quantities),  \eqref{iscat} can be written 
   \beq
    \tau_I \dot{q}^a  + q^a \approx -\hat\kappa_I h^{ab}\left(T_{\ ;b} + T\dot u_b\right) \ ,
    \eeq    
where
\beq
\tau_I = \beta_1 \hat{\kappa} T \left[ 1 +  u^c \left(\frac{\beta_1}{T} \right)_{;c}\frac{\hat{\kappa} T^2}{2}  \right]^{-1} \ ,   
\eeq
and
\beq
\hat{\kappa}_I  = \hat{\kappa} \left[ 1 +  u^c \left(\frac{\beta_1}{T} \right)_{;c}\frac{\hat{\kappa} T^2}{2}  \right]^{-1} \ .
\eeq

Now it is clear that the two equations for the heat flux are formally identical, and we can ``identify'' the parameters in the two models. 
The upshot of this is that, the models will only produce different results at higher order deviations from equilibrium. Given that this regime is hardly tested at all, 
we cannot at this stage comment on which of the two descriptions (if either) may be the most appropriate. Having said that, it is clear that the variational approach 
is formally elegant and the fact that it applies also far from equilibrium (at least in principle) may be relevant. An interesting question
concerns whether there are situations where the, rather specific, set of
higher order terms predicted by the variational analysis affect the nonlinear dynamics. 

Before we conclude the comparison, it is worth noting that the difference between the two models was apparent already from the
beginning. This is clear if we consider the stress-energy tensor. In the variational case we have
    \beq
    T_{ab}= \rho^* u_a u_b + 2 q_{(a}u_{b)} + \Psi h_{ab} + p_a \sigma_a \ .
    \eeq
Comparing this to \eqref{stress} we see that 
    \beq
    \tau = { 1 \over 3} (s^*)^2 \mathcal{B}^\s w^2
    \eeq
and
    \beq
    \tau_{ab} = 3(s^*)^2 \mathcal{B}^\s \left( w_a w_b - h_{ab} w^2 \right)
    \eeq
These terms are quadratic in $w^a$ (that is, $q^a$). Hence, it is obvious that,
in order to carry out detailed a comparison we ought to include also shear- and bulk viscosity in the model, c.f. the analysis 
of \citet{priou}.


\section{The Newtonian limit}
\label{sec.newt}

Having developed a model for heat conductivity in general relativity, and discussed the results in the context 
of the well-established Israel-Stewart model, we should also consider the implications of the model
for non-relativistic systems.  The close connection between the variational multi-fluid approach in
Newtonian gravity and extended irreversible thermodynamics has already been discussed by \citet{nilsclass}. The results 
demonstrated that a two-fluid model based on a
massive component representing the particles and a massless component representing the entropy, reproduces
many key results from the literature \citep{joubook}. In particular, the non-relativistic Cattaneo equation is obtained immediately
from the momentum conservation law for the entropy. The non-relativistic analysis is, in fact, completely analogous to the
discussion in this paper. To demonstrate this, and relate the two models, we will now work out the Newtonian, low-velocity, weak gravity, limit
of our main equations.

Let us first return to the variation of the master function $\Lambda$. Assuming low velocities we have
\beq
j^2 = -n^a s_a \approx sn \left( 1 + {w^2 \over 2c^2}\right) \ ,
\eeq
where $w^2=w_{\n\s}^2$ represents the (squared) magnitude of the relative velocity between matter and entropy ($w_{\n\s}^i=v^i_\n-v_\s^i$).
Note that we  need to keep the speed of light, $c$, explicit in this discussion.
This then leads to
\beq
dj^2 \approx s  \left( 1 + {w^2 \over 2c^2}\right) dn + n  \left( 1 + {w^2 \over 2c^2}\right) ds
+ {sn \over 2c^2} dw^2 \ ,
\eeq
and
\beq
d\Lambda \approx - \left[ n \mathcal{B}^\n +  s  \left( 1 + {w^2 \over 2c^2}\right) \mathcal{A}^{\n\s}\right]dn
- \left[ s \mathcal{B}^\s +  n  \left( 1 + {w^2 \over 2c^2}\right) \mathcal{A}^{\n\s}\right] ds - { sn \over 2c^2} \mathcal{A}^{\n\s} dw^2 \ .
\eeq

To make progress it is essential to appreciate that the Newtonian limit is singular, see for example the rigorous analysis of  \citet{cartercham}.
In order to effect a direct ``calculation'' it is useful to separate the ``ballistic'' rest-mass contribution to the master function.
That is, we use (recalling that the entropy is taken to be massless)
\beq
\Lambda = - m n c^2 - E(n,s,w^2) \ .
\eeq
From the above results it follows that
\beq
\left. { \partial \Lambda \over \partial n } \right|_{s,w^2} \approx - \left[ n \mathcal{B}^\n +  s  \left( 1 + {w^2 \over 2c^2}\right) \mathcal{A}^{\n\s}\right]  = -mc^2 - \mu \ ,
\eeq
defining the chemical potential, $\mu$,
and
\beq
\left. { \partial \Lambda \over \partial s } \right|_{n,w^2} \approx - \left[ s \mathcal{B}^\s +  n  \left( 1 + {w^2 \over 2c^2}\right) \mathcal{A}^{\n\s}\right]
= T \ ,
\eeq
which defines the temperature $T$ (as the entropy chemical potential).
Finally,
\beq
\left. { \partial \Lambda \over \partial w^2 } \right|_{n,s} \approx  - { sn \over 2c^2} \mathcal{A}^{\n\s} = - \left. { \partial E \over \partial w^2} \right|_{n,s} \equiv \alpha \ ,
\eeq
defines the Newtonian entropy entrainment parameter $\alpha$ \citep{prix}.
These three relations allow us to express $\mathcal{B}^\n$, $\mathcal{B}^\s$ and $\mathcal{A}^{\n\s}$ in terms of the Newtonian coefficients $\mu$, $T$ and $\alpha$.

We also need the weak field spacetime metric. To the required order, we have the line element
\beq
ds^2 \approx - c^2 \left( 1 + { 2 \Phi \over c^2} \right) dt^2 + g_{ij} dx^i dx^j \ ,
\eeq
where $\Phi$ is the gravitational potential and $g_{ij}$ is the flat space metric.
Meanwhile, the four velocities have components
\beq
u_\n^t = 1- {\Phi \over c^2} + {v_\n^2 \over 2c^2} \ , \qquad u_\n^i = v_\n^i \ , 
\eeq
and
\beq
u_\s^t = 1- {\Phi \over c^2} + {v_\s^2 \over 2c^2} \ , \qquad u_\s^i = v_\s^i \ .
\eeq
This means that the two fluxes become
\beq
n^a = {n u_\n^a \over c} \rightarrow \left\{ \begin{array}{ll} n^0 = {n\over c} \left(  1- {\Phi \over c^2} + {v_\n^2 \over 2c^2} \right) \ , \quad n^i = {n v^i_\n \over c} \ ,\\
n_0 = - nc \left(  1- {\Phi \over c^2} + {v_\n^2 \over 2c^2} \right)   \ , \quad n_i = {n v_i^\n \over c} \ . \end{array} \right.
\eeq

Using the different expressions in the momentum equations, keeping only terms of order unity, we obtain the Newtonian equations of motion.
For the particles we then find
\beq
n^a  \mu_{b;a}^\n \approx n (\partial_t + v_\n^j \nabla_j ) \left(  m v^\n_i + { 2 \alpha \over n} w_i^{\s\n} \right) \ ,
\eeq
\beq
n^a  \mu^\n_{a;b} \approx -n \nabla_i (\mu + m\Phi) - 2\alpha w_{\s\n}^j \nabla_i v^\n_j \ ,
\eeq
This means that the relevant momentum equation becomes
\beq
 n (\partial_t + v_\n^j \nabla_j ) \left(  m v^\n_i + { 2 \alpha \over n} w_i^{\s\n} \right) + n \nabla_i (\mu + m\Phi) + 2\alpha w_{\s\n}^j \nabla_i v^\n_j = 0 \ ,
\eeq
For the massless entropy, it is easy to see that we get
\beq
s^a  \mu_{b;a}^\s \approx s (\partial_t + v_\s^j \nabla_j ) \left(  { 2 \alpha \over s} w_i^{\n\s} \right) \ ,
\eeq
\beq
s^a  \mu^\s_{a;b} \approx -s \nabla_i T  - 2\alpha w_{\n\s}^j \nabla_i v^\s_j \ , 
\eeq
and the final momentum equation becomes
\beq
 s(\partial_t + v_\s^j \nabla_j ) \left( { 2 \alpha \over s} w_i^{\n\s} \right) + s \nabla_i T + 2\alpha w_{\n\s}^j \nabla_i v^\s_j = 0 \ .
\label{entmom}\eeq
In order to facilitate a direct comparison with the discussion by \citet{nilsclass} we now use the entropy conservation law
\beq
 s^a_{\ ;a} \approx \partial_t s + \nabla_j (s v_\s^j) = \Gamma_s \ ,
\eeq
and define
\beq
\pi_i^\s = 2\alpha w_i^{\n\s} \ .
\eeq
Then \eqref{entmom} can be written
\beq
\partial_t \pi_i^\s + \nabla_j (v_\s^j \pi_i^\s) + s \nabla_i T + \pi_j^\s \nabla_i v_\s^j = { \Gamma_\s \over s} \pi_i^\s = f_i^\s \ ,
\eeq
or
\beq
(\partial_t + v_\n^j ) \pi_i^\s - \nabla_j \left( {\pi_\s^j \pi^\s_i \over 2\alpha} \right) - \pi_\s^j \nabla_i \left( { \pi^\s_j \over 2 \alpha} \right)
+ s\nabla_i T + \pi^\s_j \nabla_i v_\n^j + \pi^\s_i \nabla_j v_\n^j =  { \Gamma_\s \over s} \pi_i^\s \ .
\eeq

We have  arrived at the Newtonian two-fluid point model that was taken as the starting point by \citet{nilsclass}. In other words, their 
Newtonian model is the natural non-relativistic counterpart to the  model developed in Section~2. That this had to be the case was, 
more or less, obvious given the discussion by \citet{livrev}, but it is still useful to have a direct comparison. After all, the derivation shows explicitly 
that the four acceleration term in \eqref{cattaneo} is a purely relativistic effect. The comparison also clarifies the physical interpretation of the 
different variables, and the meaning of the various parameters.  

\section{Discussion}

We have discussed a relativistic model for heat conduction. The model builds on the convective variational approach to multi-fluid systems
designed by \citet{carter02}, and focusses on the role of the entropy as a dynamic entity. The model assumes that the entropy can be treated as a ``fluid'' distinct from the matter flow. 
We have demonstrated how this approach leads to a relativistic version of the Cattaneo equation, encoding the thermal relaxation time that is needed to satisfy causality. 
Moreover, we have shown that the model naturally includes the non-equilibrium Gibbs relation that is a key ingredient in most approaches to extended thermodynamics. 
By focussing on the pure heat conduction problem, neglecting other ``dissipation channels'', we  compared the variational results to the celebrated second-order model
developed by \citet{IS}. The comparison showed that, despite the very different philosophies behind the two approaches, the two models are 
equivalent at first order deviations from thermal equilibrium. This was not surprising. In fact, \citet{priou} has already carried out a similar comparison of 
the corresponding second-order models (including viscosity). His results show that the two models contain the same key elements, and that they
belong to a wider class of permissible models. The simpler context of our analysis serves to clarify the main points. Finally, we worked out the non-relativistic 
limit of our results, making contact with the recent work of \cite{nilsclass}. This essentially completes the picture, and we now have a consistent framework 
for discussing causal heat conductivity in both Newtonian and relativistic dynamics. 

Our discussion obviously only scratched the surface of what is a very difficult problem. We did not address foundational issues concerning the link between this kind of
phenomenological model (e.g. the entropy ``fluid'') and the relevant microphysics/statistical physics. This is a rich and challenging area, where many 
 issues remain to be resolved, and one can imagine several interesting directions in which the current work may be developed. 
This work may also be applied in a number of exciting contexts, ranging from high-energy collisions probed by, for example, RHIC and the LHC,  to neutron star 
dynamics and issues relevant for multimessenger astronomy, and even cosmology and the evolution of the Universe itself.

\begin{acknowledgements}
As this work was  carried out, three truly inspirational colleagues passed away. Without the work done by these pioneers our understanding of relativistic 
heat problems would not be what it is today, and we would like to dedicate this work to the memories of Peter Landsberg, Bill Hiscock and J\"urgen Ehlers. 

NA acknowledges support from STFC via grant number ST/H002359/1. CSLM gratefully acknowledges support from CONACyT.

\end{acknowledgements}



\begin{thebibliography}{10}

\bibitem[{{Andersson}(2003)}]{nareview}
Andersson, N., 2003, Gravitational waves from instabilities in relativistic stars, Class. Quantum Grav. {\bf 20}, R105.


\bibitem[{{Andersson \& Comer}(2007)}]{livrev}
Andersson, N., \& Comer, G. L., 2007, Relativistic Fluid Dynamics: Physics for Many Different Scales, Living Reviews in Relativity, {\bf 10} no. 1

\bibitem[{{Andersson \& Comer}(2010)}]{nilsclass}
Andersson, N., \& Comer, G.L., 2010, Variational multi-fluid dynamics and causal heat conductivity, Proc. R. Soc. London A. {\bf 466},  1373. 

\bibitem[{{Carter}(1976)}]{carter01}
Carter, B., 1976, Regular and anomalous heat conduction: The canonical diffusion equation in relativistic thermodynamics, Journ\'ees Relativistes, Ed. M. Cahen, R. Debever, J. Geheniau, Universit\'e Libre de Bruxelles, pp 12-27.

\bibitem[{{Carter}(1988)}]{regular}
Carter, B., 1988, Conductivity with causality in relativistic hydrodynamics: the regular solution to Eckart's problem, in Highlights in gravitation and cosmology, 
Ed: B.R. iyer, A. Kembhavi, J.V. Narlikar and C.V. Vishveshwara (Cambridge Univ. Press, Cambridge 1988).

\bibitem[{{Carter}(1989)}]{carter02}
Carter, B., 1989, Covariant theory of conductivity in ideal fluid or solid media, in Relativistic Fluid Dynamics, Ed: A. Anile and M. Choquet-Bruhat, 
Springer Lecture Notes in  Mathematics vol 1385, pp 1--64.

\bibitem[{{Carter}(1991)}]{carter03}
Carter, B., 1991, Convective variational approach to relativistic thermodynamics of
  dissipative fluids, Proc. R. Soc. London. A {\bf 433}, 45.

\bibitem[{{Carter \& Chamel}(2004)}]{cartercham}
Carter, B., \& Chamel, N., 2004, Covariant Analysis of Newtonian Multi-Fluid Models for Neutron Stars
 I,  Int. J. Mod. Phys. D, {\bf 13}, 291.

\bibitem[{{Carter \& Quintana}(1972)}]{carterq}
Carter, B., \& Quintana, H., 1972, Foundations of general relativistic high-pressure elasticity theory, 
Proc. R. Soc. London A, {\bf 331}, 57.

\bibitem[{{Cattaneo}(1948)}]{cattaneo}
Cattaneo, C., 1948, Atti Seminario Univ. Modena {\bf 3}, 33.

\bibitem[{{Comer \& Joynt}(1988)}]{joynt}
Comer, G.L., \& Joynt, R., 2003, Relativistic mean field model for entrainment in general relativistic superfluid neutron stars, Phys. Rev. D {\bf 68},  023002.

\bibitem[{{Eckart}(1940)}]{eckart03}
Eckart, C., 1940, The thermodynamics of irreversible processes. iii. relativistic
  theory of the simple fluid,  Phys. Rev. {\bf 58}, 919.

\bibitem[{{Ehlers}(1973)}]{ehlers}
Ehlers, J., 1973, Survey of general relativity theory, in  Relativity, Astrophysics and Cosmology, Ed: W. Israel,
vol.~38 of {\em
  Astrophysics and Space Science Library}, pp 1--125, (Reidel, Dordrecht).

\bibitem[{{Elze, Rafelski \& Turka}(2001)}]{elze}
Elze, H.-Th., Rafelski, J., \& Turka, L., 2001, Entropy production in relativistic hydrodynamics, Phys. Lett. A {\bf 506}, 123.

\bibitem[{{Garcia-Colin \& Sandoval-Villalbazo}(2006)}]{colin01}
Garcia-Colin, L.S., \& Sandoval-Villalbazo, A., 2006, Relativistic Non-Equilibrium Thermodynamics Revisited, Journal of Non Equilibrium Thermodynamics, 
{\bf 31}, 11.

\bibitem[{{Garcia-Perciante et al}(2009)}]{colin02}
Garcia-Perciante, A.L.,  Garcia-Colin, L.S., \& and Sandoval-Villalbazo, A., 2009, 
On the nature of the so-called generic instabilities in dissipative
  relativistic hydrodynamics, Gen. Rel. Grav. {\bf 41}, 1645.

\bibitem[{{Geroch}(1995)}]{Gero}
Geroch, R., 1995, Relativistic theories of dissipative fluids, J. Math. Phys {\bf 36}, 4226.

\bibitem[{{Hiscock \& Lindblom}(1983)}]{hiscock01}
Hiscock, W.A.,\& Lindblom, L., 1983,  Stability and causality in dissipative relativistic fluids, Ann. Phys. {\bf 151}, 466.

\bibitem[{{Hiscock \& Lindblom}(1985)}]{hiscock02}
Hiscock, W.A.,\& Lindblom, L., 1985, Generic instabilities in first-order dissipative relativistic fluid theories, Phys. Rev. D {\bf 31}, 725.

\bibitem[{{Hiscock \& Lindblom}(1988)}]{hiscock03}
Hiscock, W.A.,\& Lindblom, L., 1988, Nonlinear pathologies in relativistic heat-conducting fluid theories, Phys. Lett. A {\bf 131}, 509.

\bibitem[{{Israel \& Stewart}(1979a)}]{IS2}
Israel, W., \& Stewart, J.M., 1979,  On transient relativistic thermodynamics and kinetic theory II, Proc. R. Soc. London A {\bf 365}, 43. 

\bibitem[{{Israel \& Stewart}(1979b)}]{IS}
Israel, W., \& Stewart, J.M., 1979, Transient relativistic thermodynamics and kinetic theory,  Ann. Phys. {\bf 118}, 341.

\bibitem[{{Jou et al}(1993)}]{joubook}
Jou, D., Casas-V\'azquez, J., \& Lebon, G., 1993, {\em Extended irreversible thermodynamics} (Springer, Berlin).

\bibitem[{{Landau \& Lifshitz}(1959)}]{landau}
Landau, L. D., \& Lifshitz, E. M., 1959, {\em Fluid mechanics} (Oxford, Butterworth Heinemann).

\bibitem[{{Landsberg}(1967)}]{landsberg}
Landsberg, P. T., 1967, Does a moving body appear cool?, Nature, {\bf 214}, 903.


\bibitem[{{Lindblom}(1996)}]{Lindb}
Lindblom, L., 1996, The relaxation effects in dissipative relativistic fluid theories, Ann. Phys. {\bf 247}, 1.

\bibitem[{{Muronga}(2004)}]{muronga}
Muronga, A., 2004, Causal theories of dissipative relativistic fluid dynamics for nuclear collisions, Phys. Rev. C {\bf 69}, 034903.

\bibitem[{{Olson \& Hiscock}(1990)}]{Olson}
Olson, T.S., \&  Hiscock, W.A., 1990, Stability, causality, and hyperbolicity in Carter's ``regular''
  theory of relativistic heat-conducting fluids, Phys. Rev. D {\bf 41}, 3687.

\bibitem[{{Priou}(1991)}]{priou}
Priou, D., 1991, Comparison between variational and traditional approaches to
  relativistic thermodynamics of dissipative fluids, Phys. Rev. D {\bf 43}, 1223.

\bibitem[{{Prix}(2004)}]{prix}
Prix, R., 2004, Variational description of multifluid hydrodynamics: Uncharged
  fluids, Phys. Rev. D {\bf 69}, 043001.

\bibitem[{{Samuelsson et al}(2010)}]{twostream}
Samuelsson, L., Lopez-Monsalvo, C.S., Andersson, N., \& Comer, G.L., 2010, 
Relativistic two-stream instability, Gen. Rel. Grav. {\bf 42}, 413.

\bibitem[{{Stewart}(1977)}]{Stewart}
Stewart, J.M., 1977, On transient relativistic thermodynamics and kinetic theory, Proc. R. Soc. London A {\bf 357}, 59.

\end{thebibliography}
\end{document}